\documentstyle[prl,aps,epsbox]{revtex}
\newtheorem{theorem}{Theorem}

\newtheorem{lemma}[theorem]{Lemma}

\begin{document}
\draft


\title{
Wigner-Araki-Yanase theorem on Distinguishability
}

\author{Takayuki Miyadera$\ ^*$
and Hideki Imai$\ ^{*,\dagger}$
}

\address{$\ ^*$
Research Center for Information Security (RCIS),\\
National Institute of Advanced Industrial 
Science and Technology (AIST).\\
Daibiru building 1102,
Sotokanda, Chiyoda-ku, Tokyo, 101-0021, Japan.
\\
(e-mail: miyadera-takayuki@aist.go.jp)
\\
$\ ^{\dagger}$
Graduate School of Science and Engineering Course of Electrical,
Electronic, and Communication Engineering,
\\
Chuo University. \\
1-13-27 Kasuga, Bunkyo-ku, Tokyo 112-8551, Japan .
}
%

\maketitle
\begin{abstract}
The presence of an additive conserved quantity imposes a limitation on 
the measurement process. According to the Wigner-Araki-Yanase theorem, 
the perfect repeatability and the distinguishability on the 
apparatus cannot be attained simultaneously. 
Instead of the repeatability, in this paper, 
the distinguishability on both 
systems is examined.
We derive a
trade-off inequality between the distinguishability of the 
final states on the system and the one on the apparatus. 
The inequality shows that the perfect distinguishability of
both systems cannot be attained simultaneously.
\end{abstract}
According to the Wigner-Araki-Yanase theorem, the presence of an additive conserved
quantity imposes a limitation on the measurement process. 
Wigner, and later Araki and Yanase, showed\cite{Wigner,ArakiYanase,Yanase} that in the sense of von Neumann's ideal measurement one cannot precisely measure observables which do not commute with the conserved quantity.
That is, the repeatability of the measurements and the perfect distinguishability of the final states on the measuring apparatus cannot be realized simultaneously.
On the other hand, if we abandon the repeatability condition, 
the perfect distinguishability of the final states on the apparatus can be attained\cite{GMRW,Ohira}. 
Ozawa\cite{Ozawa1,Ozawa2} has derived a quantitative relation between the noise operator and the disturbance operator by Robertson type inequality 
to discuss the trade-off between the repeatability and 
the distinguishability on the apparatus.
We, in this paper, relax the condition. We do not impose the
repeatability on the measurement process, instead 
we treat the distinguishability of the final states 
also on the system.
We ask for the quantitative trade-off between the distinguishability of the final states on the measured system and the one on the measuring apparatus.
According to our result, there is no interaction that achieves perfect distinguishability on both systems. Since our result is quantitative, it enables us to discuss the dependence on 
the size of the apparatus and the environment.
\par
Let us consider two quantum systems, a {\em system} and an {\em apparatus}.
Each system is 
described by a Hilbert space, ${\cal H}_S$ and 
${\cal H}_A$, respectively.
Suppose that there exists an additive conservative quantity.
That is, there exist an observable $L_S$ on the system and an observable
$L_A$ on the apparatus such that their summation 
$L_S+L_A$ is conserved by any physical dynamics for the closed system. 
Let us consider a pair of orthogonal vector states, 
$|\psi_0\rangle, |\psi_1\rangle\in {\cal H}_S$. 
The goal of the measurement process is to make them 
distinguishable on the apparatus by choosing an 
initial state of the apparatus and the interaction 
between the system and the apparatus.
In the case of the ideal measurement, the repeatability 
of the measurements is also imposed. 
That is, the states $|\psi_0\rangle$ and $|\psi_1\rangle$ should be invariant 
with the interaction.
We, in this paper, do not employ this repeatability condition.
We relax the condition to 
the distinguishability condition on the system.
That is, we ask if it is possible 
for the final states to be distinguishable on both systems. 
The distinguishability is characterized by 
a quantity called {\em fidelity}. 
The fidelity\cite{Uhlmann,Jozsa} between two states $\rho_0$ and $\rho_1$ is defined by 
$F(\rho_0,\rho_1):=\mbox{tr}(\sqrt{\rho_0^{1/2}\rho_1\rho_0^{1/2}})$.
It takes a nonnegative value less than $1$, and becomes smaller if the states
 are 
more distinguishable.
The perfect distinguishability corresponds to the vanishing fidelity. 
The following lemma\cite{BCFJS} justifies that the fidelity indeed represents 
the distinguishability.
\begin{lemma}\label{lemma1}
The fidelity equals the square root of 
minimum overlap coefficient between 
two probability distributions $p_0$ and $p_1$:
\begin{eqnarray*}
F(\rho_0,\rho_1)=\min_{\{E_{\alpha}\}:POVM}\sum_{\alpha}
\sqrt{p_0(\alpha)p_1(\alpha)},
\end{eqnarray*} 
where $p_0$ and $p_1$ are defined by 
$p_0(\alpha)=\mbox{tr}(\rho_0 E_{\alpha})$ 
and $p_1(\alpha)=\mbox{tr}(\rho_1 E_{\alpha})$. 
The minimum is taken over all the possible positive 
operator valued measures (POVMs), where a POVM $\{E_{\alpha}\}$ is 
a family of the positive operators satisfying 
$\sum_{\alpha}E_{\alpha}={\bf 1}$. 
Moreover, the minimum is attained by a projection 
valued measure (PVM), where a PVM $\{E_{\alpha}\}$
is a family of the projection operators satisfying 
$\sum_{\alpha}E_{\alpha}={\bf 1}$.
\end{lemma}
\par
This lemma plays an essential role in the proof of our theorem.
In the presence of 
the additive conserved quantity, we have the following theorem.
\begin{theorem}\label{maintheorem}
As described above, let us consider a pair of orthogonal states
$|\psi_0\rangle, |\psi_1\rangle \in {\cal H}_S$ in the presence of 
the additive conserved quantity, $L_S+L_A$.
For any initial state $\sigma$ on the apparatus and the 
unitary dynamics $U$ satisfying conservation law, the final states 
$\rho_0:=U(|\psi_0\rangle \langle \psi_0| \otimes \sigma)U^*$ 
and $\rho_1:=U(|\psi_1\rangle \langle \psi_1| \otimes \sigma)U^*$ 
satisfy the following:
\begin{eqnarray}
|\langle \psi_0|L_S|\psi_1\rangle|
\leq \Vert L_A\Vert F(\rho_0^{S},\rho_1^S)
+\Vert L_S \Vert F(\rho_0^{A},\rho_1^{A}),
\label{tradeoff}
\end{eqnarray}
where $\rho_i^{S}$ is the final state $\rho_i$ restricted 
to the system and $\rho_i^A$ is the one restricted 
to the apparatus, and $F(\cdot,\cdot)$ is the fidelity,
and $\Vert \cdot \Vert$ represents the operator norm defined 
as $\Vert A\Vert :=\sup_{\varphi\neq 0,\varphi \in {\cal H}}
\frac{\Vert A\varphi\Vert}{\Vert \varphi\Vert}$ for any operator 
$A$ on a Hilbert space ${\cal H}$.
\end{theorem}
{\bf Proof:}
By the purification of $\sigma$, we obtain 
a dilated Hilbert space and a vector state for the apparatus. 
We write the dilated Hilbert space as ${\cal H}_A$ for simplicity 
and the vector state as $|\Omega\rangle$. The dilated unitary 
operator $U \otimes {\bf 1}$ is also abbreviated as $U$.
Let us define the initial vector states $|\Psi_i \rangle
:=|\psi_i\rangle \otimes |\Omega \rangle$ for $i=0,1$. 
As Wigner-Araki-Yanase's original discussion, we consider 
the following quantity:
\begin{eqnarray}
\langle \psi_0|L_S|\psi_1\rangle&=&
\langle \Psi_0|L_S+L_A|\Psi_1\rangle
\nonumber \\
&=&\langle \Psi_0|U^*(L_S+L_A)U|\Psi_1\rangle
\nonumber \\
&=&\langle \Psi_0|U^*L_SU|\Psi_1\rangle
+\langle \Psi_0|U^*L_AU|\Psi_1\rangle,
\label{fundamental}
\end{eqnarray}
where in the first line we have used $\langle \Psi_0|L_A|\Psi_1\rangle
=\langle \psi_0|\psi_1\rangle\langle \Omega |L_A|\Omega\rangle =0$.
Now we consider an arbitrary projection 
valued measure (PVM) $\{E_{\alpha}\}$ on the system 
and an arbitrary PVM $\{P_j\}$ on the apparatus.
Since $\sum_{\alpha}E_{\alpha}=\sum_{j}P_j={\bf 1}$ holds, 
the right hand side of (\ref{fundamental}) can be written as 
$\sum_{j} \langle \Psi_0|U^*P_j L_SU|\Psi_1\rangle
+\sum_{\alpha} \langle \Psi_0|U^*E_{\alpha} L_AU|\Psi_1\rangle$.
By using commutativity $[P_j,L_S]=[E_{\alpha},L_A]=0$, 
we obtain,
\begin{eqnarray*}
\langle \psi_0|L_S|\psi_1\rangle&=&
\sum_{j}\langle \Psi_0|U^*P_j L_S P_j U|\Psi_1\rangle
+
\sum_{\alpha} \langle \Psi_0|U^*E_{\alpha} L_A E_{\alpha}
U|\Psi_1\rangle.
\end{eqnarray*}
Taking absolute value of the both sides, we obtain, 
\begin{eqnarray*}
|\langle \psi_0|L_S|\psi_1\rangle|&\leq &
\sum_{j}|\langle \Psi_0|U^*P_j L_S P_j U|\Psi_1\rangle|
+\sum_{\alpha}|\langle \Psi_0|U^*E_{\alpha} L_A E_{\alpha}
U|\Psi_1\rangle|
\\
&\leq& \Vert L_S\Vert 
\sum_j \sqrt{\langle \Psi_0|U^*P_j U|\Psi_0\rangle
\langle \Psi_1|U^*P_j U|\Psi_1\rangle}
\\
&+&
\Vert L_A\Vert \sum_{\alpha}\sqrt{
\langle \Psi_0|U^*E_{\alpha} U|\Psi_0\rangle
\langle \Psi_1 |U^* E_{\alpha} U|\Psi_1\rangle}.
\end{eqnarray*}
We here choose the particular PVMs, $\{E_{\alpha}\}$ and $\{P_j\}$, which 
attain the fidelity. Thanks to the lemma\ref{lemma1}, we finally obtain,
\begin{eqnarray*}
|\langle \psi_0|L_S|\psi_1\rangle|\leq
\Vert L_A\Vert F(\rho_0^{S},\rho_1^{S})
+\Vert L_S\Vert F(\rho_0^A,\rho_1^A).
\end{eqnarray*}
It ends the proof.
\hfill Q.E.D.
\par
According to this theorem, we obtain the following theorem.
\begin{theorem}
Under the setting of the theorem\ref{maintheorem},
the perfect 
distinguishability for both systems cannot be 
attained simultaneously.
\end{theorem}
{\bf Proof:}\\
The vanishing fidelities in (\ref{tradeoff}) 
contradict with the nonvanishing left hand side.
\hfill Q.E.D.
\par
Let us consider the simplest example. The system is a spin $1/2$ system. 
The conserved quantity is the 
$z$-component of the spin, $S_z+L_A$, 
where $L_A$ is the $z$-component of the spin 
on the apparatus. $S_z$ is written with the 
eigenvectors, $|1\rangle$ and $|-1\rangle$, as 
$S_z=\frac{\hbar}{2}(|1\rangle\langle 1|-|-1\rangle \langle -1|)$.
The observable to be measured $S_k$ is a component of spin in 
another direction.
That is, 
the states to be distinguished by the measurement process are
$|\psi_1\rangle:=\alpha|1\rangle+\beta|-1\rangle$ and 
$|\psi_0\rangle:=\overline{\beta}|1\rangle-\overline{\alpha}|-1\rangle$, 
where $|\alpha|^2+|\beta|^2=1$ with 
$\alpha \neq 0, \beta \neq 0$. 
The observables $S_z$ and $S_k$ do not 
commute with each other. In fact, 
\begin{eqnarray*}
\langle \psi_0|S_z|\psi_1\rangle =\hbar \alpha \beta
\end{eqnarray*}
holds. If we assume the rigorous repeatability as in the original 
Wigner-Araki-Yanase theorem,
the state change for the dilated Hilbert space should be 
written as,
\begin{eqnarray*}
|\psi_j\rangle \otimes |\Omega\rangle
\mapsto |\psi_j\rangle \otimes |\phi_j\rangle
\end{eqnarray*}
for $j=0,1$. It gives,
\begin{eqnarray*}
\langle \psi_0|S_z|\psi_1\rangle
=\langle \psi_0|S_z|\psi_1\rangle 
\langle \phi_0|\phi_1\rangle,
\end{eqnarray*}
and thus $|\phi_0\rangle =|\phi_1\rangle$ holds.
Therefore there is no distinguishability on 
the apparatus side in this case.
On the other hand, if we do not impose the repeatability, 
the distinguishability on both systems is 
partially attained. In particular, even the perfect 
distinguishability on the apparatus allows 
the partial distinguishability on the system. 
Ohira and Pearle\cite{Ohira} has constructed the following interaction
between the system and the spin $1/2$ apparatus:
\begin{eqnarray*}
|\psi_1\rangle \otimes \sqrt{\frac{1}{2}}(|1\rangle +|-1\rangle)
&\mapsto&
(\alpha |1\rangle +\beta |-1\rangle)
\otimes \sqrt{\frac{1}{2}}(|1\rangle +|-1\rangle)
\\
|\psi_0\rangle \otimes \sqrt{\frac{1}{2}}(|1\rangle +|-1\rangle)
&\mapsto&
(\overline{\beta} |1\rangle +\overline{\alpha} |-1\rangle)
\otimes \sqrt{\frac{1}{2}}(|1\rangle -|-1\rangle).
\end{eqnarray*}
It gives the fidelity $F(\rho_0^A,\rho_1^A)=0$ and 
$F(\rho_0^S,\rho_1^S)=2|\alpha \beta|$. 
Since $\Vert L_A\Vert=\hbar/2$ holds, this interaction satisfies,
\begin{eqnarray*}
|\langle \psi_0|S_z|\psi_1\rangle|
=\Vert L_A\Vert F(\rho_0^S,\rho_1^S),
\end{eqnarray*}
which is the equality version of our theorem.
\par
In the following we consider the effect of the environment.
We treat a tripartite system which consists of 
the system, the apparatus, and the environment.
The Hilbert space of the environment is written as 
${\cal H}_E$. 
On the environment an operator $L_E$ is defined, and 
the conserved quantity is $L_S+L_A+L_E$. We divide the whole system 
into ${\cal H}_S$ and ${\cal H}_A\otimes {\cal H}_E$.
Application of the theorem\ref{maintheorem} to it derives,
\begin{eqnarray*}
|\langle \psi_0|L_S|\psi_1\rangle|
\leq (\Vert L_A \Vert +\Vert L_E\Vert)
F(\rho_0^S,\rho_1^S)+\Vert L_S\Vert F(\rho_0^{AE},\rho_1^{AE}),
\end{eqnarray*}
where $\rho_j^{AE}$ is a state over the apparatus and the environment. 
Since the partial trace does not reduce the fidelity\cite{BCFJS}, 
we obtain,
\begin{eqnarray*}
|\langle \psi_0|L_S|\psi_1\rangle|
\leq (\Vert L_A \Vert +\Vert L_E\Vert)
F(\rho_0^S,\rho_1^S)+\Vert L_S\Vert F(\rho_0^{A},\rho_1^{A}).
\end{eqnarray*}
This inequality shows that to attain high distinguishability 
on both systems simultaneously the large apparatus or environment 
is necessary. 
\\
{\bf Acknowledgments:}
T.M. thanks H.Moriya for valuable comments.

\end{document}